# The signature of microlensing in QSO variability–redshift correlations

Tal Alexander
*School of Physics and Astronomy, Tel-Aviv University, Tel-Aviv 69978, Israel.*

**ABSTRACT**

A recently discovered inverse correlation between QSO redshift and long-term continuum variability timescales was suggested to be the signature of microlensing on cosmological scales (Hawkins 1993). A general theoretical method for calculating such correlations is presented and applied to various lensing scenarios in the framework of $\Lambda = 0$ Friedmann cosmologies. It is shown that the observed timescales can be strongly influenced by the observational limitations: the finite duration of the monitoring campaign and the finite photometric sensitivity. In most scenarios the timescales increase with source redshift, $z_s$, although slower than the $1 + z_s$ time dilation expected of intrinsic variability. A decrease can be obtained for an extended source observed with moderate sensitivity. In this case, only lenses no further away than several hundreds Mpc participate in the lensing. The resulting optical depth is too small to explain the common long-term QSO variability unless an extremely high local lens density is assumed. These results do not support the idea that the reported inverse correlation can be attributed to microlensing of a uniform QSO sample by a uniform distribution of lenses. The possibility of using observations at various wavelengths and QSO samples at various positions to identify microlensing in QSO variability is also discussed.

**Key words:** gravitational lensing – quasars:general – dark matter

## 1 INTRODUCTION

The possibility that long-term QSO continuum variability carries the signature of microlensing bears upon two important issues: It may yield information on the amount and nature of compact dark matter on cosmological scales. It may also be used to investigate intrinsic properties of QSO, such as the size of the accretion disk (Rauch & Blandford 1991) and the origin of the ubiquitous long-term variability, which is yet to be satisfactorily explained in terms of an internal QSO mechanism (Rees 1984).

A basic problem in identifying this signature is that the microlensing-induced variability, if indeed it exists, may be masked by intrinsic variability on similar timescales. This problem can be addressed by a two step procedure. The first is to assume that the variability is due exclusively to microlensing, choose a measurable property of the light curves, such as max $\Delta m$, the maximal amplitude difference, and theoretically predict its behavior under this assumption. The second step is to use the observed light curves to obtain *upper bounds* on the effect of microlensing on the property under study. These, in turn, can be translated into bounds on physical parameters such as the lens mass, $\Omega_\ell$, the critical density fraction in lenses, and the size of the accretion disk.

Such a program was carried out by Schneider (1993), who demonstrated by numeric simulations that microlensing in a flat Friedmann universe is expected to induce more high max $\Delta m$ light curves than are actually observed, and thus obtained upper limits on $\Omega_\ell$ in various lens mass ranges.

The mean variability timescale is another measurable property of the light curve. Hawkins (1993) estimated the variability timescale in a sample of about 300 QSO monitored over 17 years by the width of the zero time-lag peak in the light curve autocorrelation function (ACF). He reported a systematic *decrease* of the typical timescales with increasing source redshift, $z_s$, contrary to the $1 + z_s$ time dilation expected for intrinsic variations in a uniform QSO sample. Hawkins suggested that this anti-correlation between variability and source redshift is the signature of microlensing, but did not present detailed calculations to support this idea. The theoretical $z_s$ dependence of microlensing timescales was briefly discussed by Canizares (1982), who calculated the rise time for amplification by a given factor in the 'empty cone' cosmological model, taking into account corrections for amplification by more than one lens. The general form of the timescale's $z_s$ dependence was, however, not reported and the few numerical values quoted, indicating an increase of the timescale with $z_s$, were for a limited $z_s$ range and for the $q_0 = 1/2$ case only (the $q_0$ dependence was reported to be weak).

This work is a theoretical study of the correlation between the variability timescale and the source redshift (henceforth $\tau$–$z$ correlations) under the hypothesis that long-term QSO variability over periods of years is dominated by microlensing. A general approximate method, independent of the details of the source surface brightness distribution, is developed in section 2 for calculating the $\tau$–$z$ correlations in $\Lambda = 0$ Friedmann cosmologies. The resulting closed expression is validated against exact results for constant surface brightness sources. The predictions for the $\tau$–$z$ correlations are presented and compared to the Hawkins results in section 3. Additional methods for detecting microlensing in $\tau$–$z$ correlations are suggested. The sensitivity of the results to the observational procedure, the assumptions required for interpreting the Hawkins results as evidence for microlensing and the limitations of the method presented here, are discussed in section 4.

## 2 CALCULATIONS

The $\tau$–$z$ correlations will be calculated with the assumption of a simple model whereby the typical source radius, $\ell_s$, and the mean transverse source velocity are independent of the source redshift, $z_s$, and that the typical lens mass, $m_\ell$, and the mean transverse lens velocity are both independent of the lens redshift, $z_\ell$. It is assumed that the lenses are compact objects, that is contained within their Einstein radius, $\ell_e$ (equation A5), and that they are homogeneously distributed with constant comoving density in a matter-dominated, $\Lambda = 0$ Friedmann universe.

An additional assumption, the shape of the amplified peak, is required for calculating the timescale, whether defined by an amplification threshold, by the mean FWHM or any other criterion. When the source is point-like, the amplification is a known analytic function of the lens mass and impact parameter only (Paczyński 1986). Realistic sources, however, are not point-like and furthermore, even if this is a good approximation for $z_\ell \ll z_s$, it inevitably breaks down as $z_\ell$ approaches $z_s$ (See figure 2 below). For an extended source, the amplified light curve depends also on the source's surface brightness distribution and on the ratio between the source's angular size and the lens cross-section. As this ratio increases with $z_\ell$, a larger fraction of the observed flux is unlensed and the maximal amplification (ratio of lensed to unlensed flux) drops to one. Rather than introduce one specific, but poorly motivated, surface brightness model, an approximate geometrical method is developed below which assumes only the existence of a maximal cutoff on $z_\ell$ beyond which the magnification is negligible. While the lensing timescale thus defined cannot be directly related to an amplification threshold, the advantages of this method lie in its generality and in the simplicity of the geometrical interpretation and of the resulting expressions. Given a surface brightness model, an exact expression for the amplification threshold timescale can be derived from the approximate one by a simple modification. It will be shown that the exact results thus obtained for a constant surface brightness model display the same general trends as those obtained by the approximate method. It will therefore be argued that the approximate method is adequate for investigating the overall properties of the $\tau$–$z$ correlations and that the conclusions depend only weakly on the details of the source emission.

We will assume that at any given time the observed QSO is undergoing amplification by no more than a single lens, i.e. that the optical depth for microlensing is small. In this case the mean timescale of an amplified peak is given by

$$\langle \tau(z_s) \rangle = \int_0^{z_s} \tau(z_\ell) \frac{dP}{dz_\ell} dz_\ell \Big/ \int_0^{z_s} \frac{dP}{dz_\ell} dz_\ell, \quad (1)$$

where $\tau(z_\ell)$ is the typical lensing timescale for a lens at redshift $z_\ell$ and $dP/dz_\ell$ the differential probability for finding a lens near the line of sight, at $z_\ell$. The small optical depth implies that the typical time between the light curve peaks is much larger than their width. $\langle \tau(z_s) \rangle$ can therefore be interpreted as the width of the ACF zero time-lag peak, which measures in this case the width of a typical peak.

We begin by considering the case where the observer and source are stationary and only the lens is moving. In the simple case of a point source, $\tau$ is defined as the crossing time of $\pi \ell_e / 2$, where the $\pi/2$ factor results from averaging over all impact parameters from 0 to $\ell_e$. In the case of an extended source (Fig. 1a), the Einstein radius $\ell_e$ is generalized to $\ell_\ell$, the distance between the two opposed points along the lens trajectory through the area subtended by the source which are no further away than $\ell_e$ from the area's boundary. The geometry describing a stationary lens and source and a moving observer is presented in figure 1b, in which case $\tau$ is defined as the crossing time of $\pi \ell_o / 2$.

In the general case the lensing timescale is determined by the combined motion of the observer, lens and source. The effective two dimensional apparent motion as measured by the observer is

$$\vec{\mu}_{\text{eff}} = \frac{\vec{v}_{\perp \ell}}{(1+z_\ell) D_{o\ell}} - \frac{\vec{v}_{\perp s}}{(1+z_s) D_{os}} - \frac{\ell_\ell \vec{v}_{\perp o}}{\ell_o D_{o\ell}} \quad (2)$$

where $D_{o\ell}$ and $D_{os}$ are the observer–lens and observer–source angular diameter distances, respectively, and $v_{\perp o}$ the observer's transverse velocity. The contribution of $v_{\perp o}$ to the apparent motion is obtained by noting that by the time the observer crosses $\ell_o$, the lens appears to cross $\ell_\ell$[*]. The effective apparent motion, $\mu_{\text{eff}}$, can be estimated by $\langle \vec{\mu}_{\text{eff}}^2 \rangle^{1/2}$ with the mean taken over the 3D velocity distribution functions of the lenses and sources and over the $v_{\perp o}$ values of the QSO in the $z_s$ interval. Assuming that the distribution functions of the lenses and sources are independently Maxwellian with 2D r.m.s velocities of $u_\ell$ and $u_s$, respectively, we obtain

$$\mu_{\text{eff}}(z_\ell, z_s) = \left\{ \left( \frac{u_\ell}{(1+z_\ell) D_{o\ell}} \right)^2 + \left( \frac{u_s}{(1+z_s) D_{os}} \right)^2 + \left( \frac{\ell_\ell}{\ell_o D_{o\ell}} \right)^2 \langle v_{\perp o}^2 \rangle \right\}^{1/2}, \quad (3)$$

where $u$ is related to the velocity dispersion $\sigma$ by $u = \sqrt{2}\sigma$. For a survey over a small angular area of the sky, $\langle v_{\perp o}^2 \rangle = (v_o \sin \alpha)^2$, where $\alpha$ is the angle between the line of sight and $\vec{v}_o$, the observer's peculiar motion relative to the cosmic microwave background (CMB). For an all-sky survey, $\langle v_{\perp o}^2 \rangle = \frac{2}{3} v_o^2$. The lensing timescale is

$$\tau(z_\ell, z_s) = \frac{\pi \ell_\ell}{2 D_{o\ell} \mu_{\text{eff}}}. \qquad (4)$$

---

[*] Eq. 2 is a generalization of eq. B9 in Kayser et al (1986) which applies only to zero-width caustics or to point-like lenses with $\ell_e \ll \ell_s$. (The case of setting $r_0 = r_\ell$ in fig. 1b.)

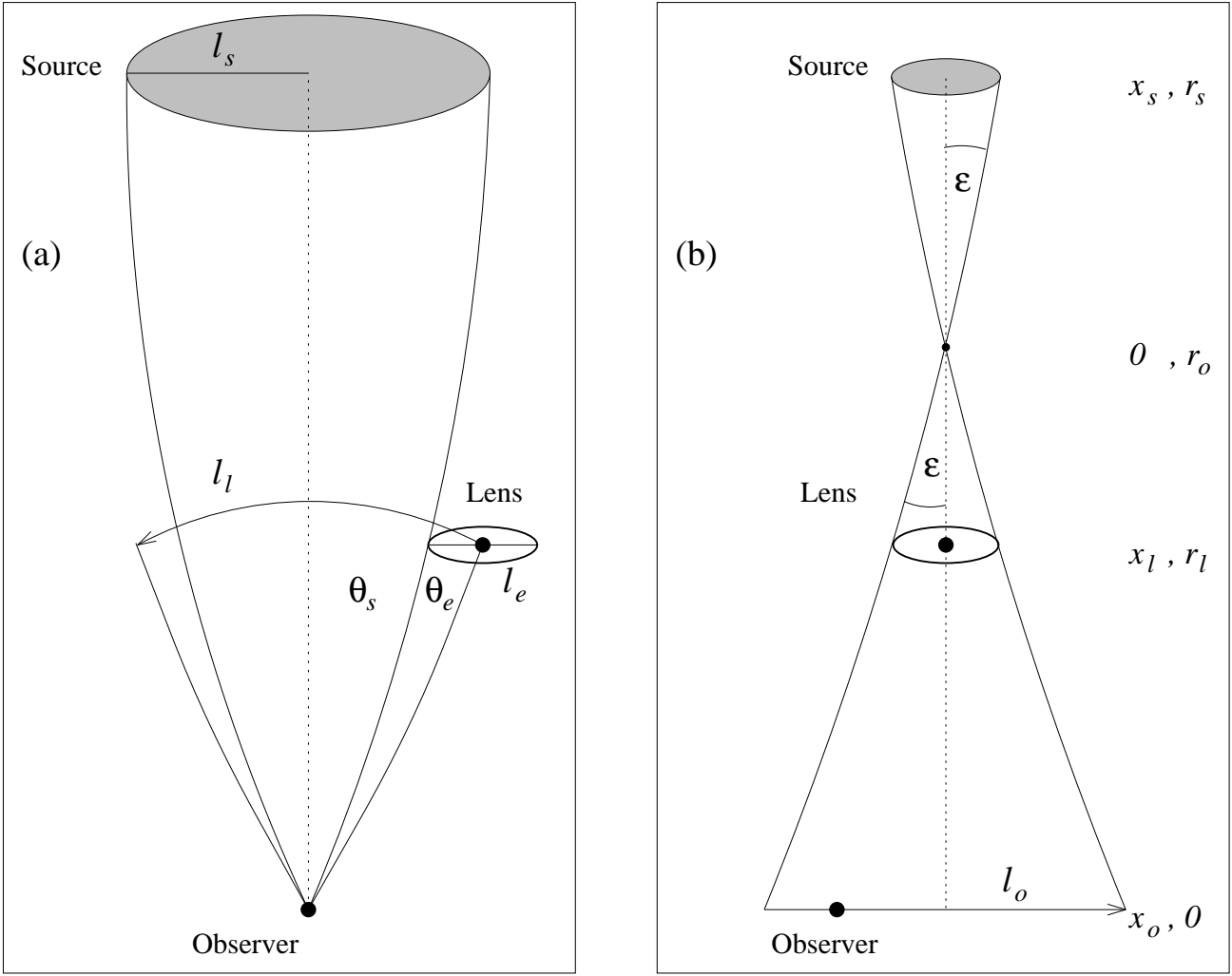

**Figure 1.** The geometry of lensing by an extended source. The optical axis (dotted line) is the geodesic connecting the observer and source center at maximum amplification. a) Moving lens and source in a frame where the observer is stationary. b) Moving observer in a frame where the lens and source are stationary. $r$ is the radial coordinate in a frame centered on the observer and $x$ is the radial coordinate in a frame centered on the intersection of the two limiting rays (see appendix).

The differential lensing probability is obtained by integrating over the lenses in the volume element subtended by $\ell_\ell$,

$$dP = \rho_\ell c |dt| 2\pi R_\ell^2 r_\ell^2 \int_0^{\theta_s + \theta_e} \sin\theta \, d\theta, \qquad (5)$$

where $\rho_\ell$ is the space density of the lenses, $R_\ell$ the expansion factor at the deflection event, $r_\ell$ the coordinate distance to the lens and $\theta_e$, $\theta_s$ the angular diameters of the Einstein ring and the source, respectively. In the small angle approximation $dP$ is the sum of three terms

$$dP = \rho_\ell \left\{ \pi \ell_e^2 c |dt| + 2\pi \frac{D_{o\ell}}{D_{os}} \ell_e \ell_s c |dt| + dV \right\}, \qquad (6)$$

where $dV$ is the differential volume subtended by the source[†]. The lensing path length $\ell_\ell$ was defined so as to obtain the required limiting behavior of $dP$. In the limit of a point source $dP$ reduces to the differential optical depth

$$dP = \rho_\ell \pi \ell_e^2 c |dt|, \qquad (7)$$

and the definition of the lensing time coincides with that of an $A_{\min} = 1.34$ amplification threshold. In the limit of a very extended source (negligible $\ell_e$), $dP$ reduces to the differential number of lenses in the volume subtended by the source

$$dP = \rho_\ell dV. \qquad (8)$$

Up to this point it was assumed that the source amplification is caused by the motion of a single compact object across the line of sight. In the case of an extended source,

---

[†] Eq. 6 is a generalized, differential form of the optical depth given in Kochanek & Lawrence (1990)

where $\theta_s/\theta_e \gg 1$, the amplification by a single compact object becomes very small. Refsdal & Stabell (1991) suggest that even when $\theta_s/\theta_e$ is as high as 10, amplifications of up to 0.1m can be caused by a localized population of lenses, for example in a galactic halo. In this case the variability is due to the statistical Poissonian fluctuation in the number of lenses in the solid angle subtended by the source. The lensing timescale formalism presented here can address this type of microlensing variability without further modification, since the typical timescale for the Poissonian fluctuations is of the order of the crossing time of $\ell_\ell$ and the small optical depth assumption remains valid since the whole lens population is associated with a single redshift value.

It is convenient to transform to the dimensionless quantities:

$$\tilde{D} = \frac{D}{R_{\rm H}},$$
$$\tilde{\ell} = \frac{\ell}{\sqrt{\frac{4Gm_\ell}{cH_0}}},$$
$$\tilde{u} = \frac{u}{v_o},$$
$$\tilde{\mu} = \frac{\mu}{\left(\frac{v_o}{R_{\rm H}}\right)},$$
$$\tilde{\tau} = \frac{\tau}{\left(\sqrt{\frac{4Gm_\ell}{cH_0}}\bigg/v_o\right)}, \qquad (9)$$

where $H_0$ the Hubble constant and $R_{\rm H} = c/H_0$ the Hubble radius. Using the expressions given in the appendix, all the relevant quantities can be expressed in terms of $q_0$, the deceleration parameter, $z_\ell$, $z_s$ and $\tilde{\ell}_s$, which measures the deviation from the case of a point source. We obtain

$$dP = \frac{3}{2}\Omega_\ell \left\{ \left(\frac{\tilde{D}_{o\ell}\tilde{D}_{\ell s}}{\tilde{D}_{os}}\right) + \tilde{\ell}_s \left(2\tilde{D}_{\ell s}^{1/2}\left(\frac{\tilde{D}_{o\ell}}{\tilde{D}_{os}}\right)^{3/2}\right) + \tilde{\ell}_s^2 \left(\frac{\tilde{D}_{o\ell}}{\tilde{D}_{os}}\right)^2 \right\} \frac{1+z_\ell}{\sqrt{1+2q_0 z_\ell}} dz_\ell, \qquad (10)$$

where $\tilde{D}_{\ell s}$ the dimensionless lens–source angular diameter distance, and

$$\tilde{\tau} = \frac{\pi}{2}\frac{\tilde{\ell}_\ell}{\tilde{D}_{o\ell}}\left\{\left(\frac{\tilde{u}_\ell}{(1+z_\ell)\tilde{D}_{o\ell}}\right)^2 + \left(\frac{\tilde{u}_s}{(1+z_s)\tilde{D}_{os}}\right)^2 + \left(\frac{\tilde{\ell}_\ell}{\tilde{\ell}_o\tilde{D}_{o\ell}}\right)^2 \sin^2\alpha \right\}^{-1/2}, \qquad (11)$$

where it is assumed that the QSO survey covers a small angular area and that $u_\ell$ and $u_s$ do not depend on $z_\ell$ and $z_s$. Finally, equations 10 and 11 are substituted into equation 1 and the lensing timescale integral is calculated numerically.

The relative angular size of the source

$$\frac{\theta_s}{\theta_e} = \tilde{\ell}_s \sqrt{\frac{\tilde{D}_{o\ell}}{\tilde{D}_{\ell s}\tilde{D}_{os}}} \qquad (12)$$

diverges as $z_\ell$ approaches $z_s$ when $\tilde{\ell}_s > 0$ (Fig. 2) and therefore there exists a maximal lens redshift for detectable amplification. This can be handled by introducing a maximal angular size ratio cutoff $C_\theta$, which is translated to a cutoff on the upper integration limit in eq. 1. Different values of $C_\theta$ are expected to approximate the behavior of different surface brightness models. $C_\theta$ will be considered a free parameter of the calculations.

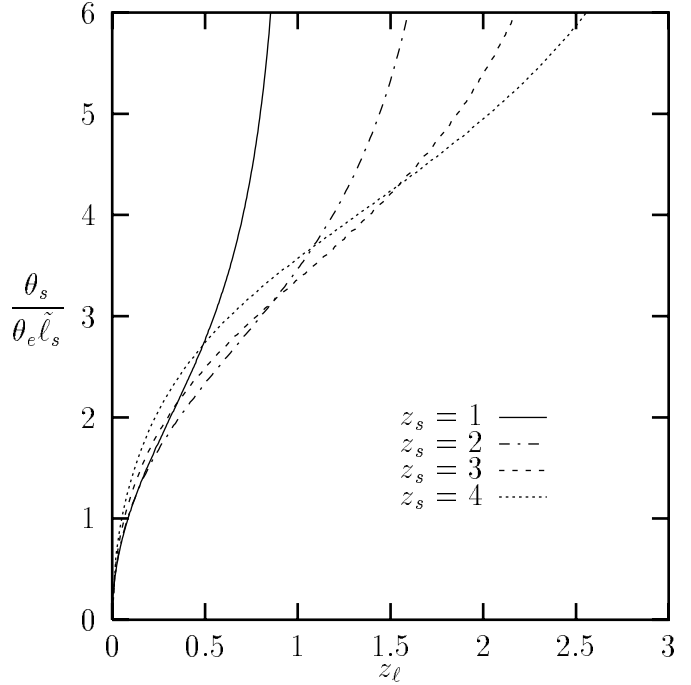

**Figure 2.** the source to lens angular diameter ratio $\theta_s/\theta_e/\tilde{\ell}_s$ for $q_0 = 1/2$ and $z_s = 1$, 2, 3 and 4. The ratio diverges as $z_\ell$ approaches $z_s$. Note that for small lens redshifts this ratio is *bigger* for the more distant QSO.

The finite duration of the observational campaign, $T$, sets an upper limit on the detectable variation timescales. Variations on scales of $\sim T$ or longer will appear as constant contributions to the light curve and therefore any measure of the light curve variability will reflect only the contribution of lenses from regions where $\tau < T$. This requires a modification of the integration limits in eq. 1, which is determined by introducing a maximal timescale cutoff $C_\tau$. Figure 3 displays the $z$ dependence of $\tau$ for the three cases where only one of the observer, lens or source is moving and for a case of combined motion. An upper cutoff on $z_\ell$ is required when the effective motion is due mainly to the observer and a lower cutoff when it is mainly due to the source motion. When the motion is mainly due to the lens, $\tau$ has a maximum and this may result in the exclusion of an intermediate $z_\ell$ range from the integration interval. The lensing timescales due to the lens motion are significantly shorter than those due to the observer or source motion. In the case of combined motion, when $v_{\perp o}$ is comparable to $u_\ell$ and $u_s$, the lens motion dominates the overall behavior of $\tilde{\tau}$.

The exact amplification, $A(\theta_s/\theta_e, \theta_\ell/\theta_e)$, can be calculated when the source surface brightness model is given explicitly, and inverted to obtain $\theta_\ell$ as a function of a minimal amplification $A_{\min}$. The amplification threshold timescale can be obtained from the approximate expressions above by

### 3.1 Fitting the Hawkins $\tau$–$z$ correlations

The QSO sample analyzed by Hawkins (1993) was taken from an area of 19 square degrees at right ascension 21.5h and declination $-45°$ (i.e. a small angular area survey with $\alpha = 125°$) that was monitored once a year over 17 years. About 1000 QSO candidates were initially selected by variability and about 300 of those were included in the sample. The QSO were divided into two luminosity bins, $-26 < M < -25$ and $-27 < M < -26$, and into two redshift bins, $1 < z_s < 2$ and $2 < z_s < 3$. The timescale was estimated by fitting the mean ACF of each of the 4 bins to the function $b(\exp(-at) - 1) + 1$ where $a$ and $b$ are free parameters and $\tau = \ln(2)/a$. Hawkins reports that for the low luminosity QSO the light curves have mean timescales of 2.3 and 1.4 yr for the low and high redshift bins respectively. For the high luminosity QSO the mean timescales are 5.2 and 3.5 yr respectively. In both luminosity bins the timescale *decreases* by a factor of 0.7 as $\bar{z}_s$ increases from 1.5 to 2.5.

In applying the theoretical results derived above to this data, It will be assumed that $\ell_s$ is independent of $z_s$ within the 1m-wide bins of absolute magnitude and that the source is of an approximately uniform color. The implications of deviations from these assumptions are discussed in section 4.

Figures 4a and 4b compare the calculated $\tau$–$z$ correlations for the Hawkins sample parameters with the observed ones for $\sqrt{h/m_1} = 3$ and 10 and a nearly point-like source with $\tilde{\ell}_s = 0.05$, corresponding to a black hole mass of $M_8 = 1.7$ and 0.5, respectively. The flattening of the predicted lensing timescale is due to the onset of the maximal time cutoff $C_\tau$. As $\sqrt{h/m_1}$ increases, the typical lensing timescales become shorter and the cutoff comes into effect at higher $z_s$. The observed $\tau$–$z$ correlations differ from the calculated ones in their negative slope. The predicted timescales can be shortened by assuming higher lens and source velocities, but this results in a case similar to that shown in figure 4b, where the theory predicts $\tau$ to be a rising function of $z_s$, contrary to what is observed.

Figures 4c and 4d compare the lensing probability in the two cases. The onset of the maximal time cutoff excludes lenses at intermediate redshifts and decreases the lensing probability for $\sqrt{h/m_1} = 3$ relative to that of $\sqrt{h/m_1} = 1$. It also introduces a peak in the lensing probability at a $z_s$ value that increases with $q_0$. The $\Omega_\ell$ values required for $P \sim 0.3$, which is consistent with the assumption that every QSO is occasionally undergoing lensing, is of the order of 1 for both $\sqrt{h/m_1} = 1$ and $\sqrt{h/m_1} = 3$.

A decrease in $\tau$, such as seen in the observations, can be obtained for an extended source and a small angular size cutoff. Figure 5a compares the Hawkins results to a case of $\sqrt{h/m_1} = 10$ and $\tilde{\ell}_s = 1$ (corresponding for example to $m_1 = 0.01$, $h = 1$ and $M_8 = 10$). The decrease in the slope of the predicted timescales is caused by the small $C_\theta = 1$ cutoff. For $q_0 > 0$ and small enough $z_\ell$ values, $\theta_s/\theta_e$ increases with increasing $z_s$, thereby causing a *decrease* of the maximal allowed $z_\ell$, that is, the more distant QSO are lensed by nearer lenses (Fig. 2). In this case the observations and the predicted timescales show a qualitative agreement for $q_0 \geq 1/4$. The maximal $z_\ell$ values, as determined by the angular size cutoff, are 0.12, 0.09, and 0.06 for the cases of $q_0 = 1/4$, 1/2 and 1, respectively, implying that the microlensing objects are only several hundred Mpc away regardless of the source

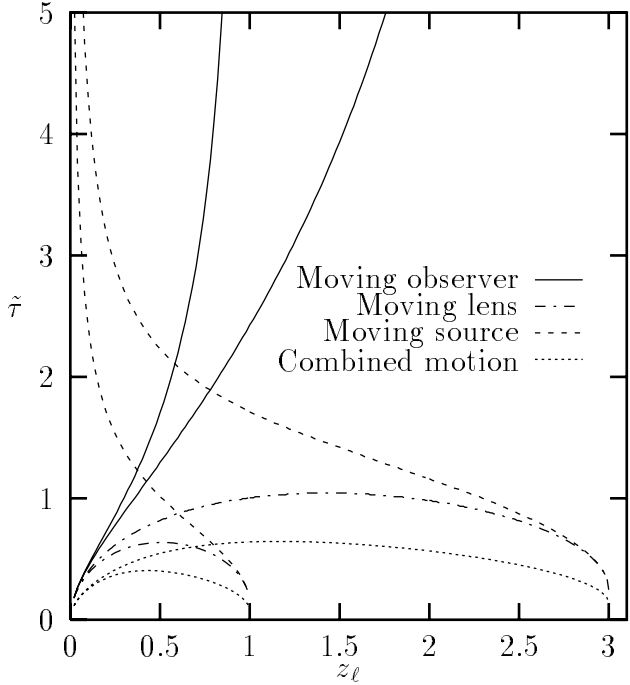

**Figure 3.** The lensing timescale for motion of observer only, lens only, source only and combined motion of all three as function of $z_\ell$ and $z_s$ for $q_0 = 1/2$ and $z_s = 1$ and 3. $\tilde{\ell}_s = 0.05$, $\alpha = 125°$ and $\tilde{u}_\ell = \tilde{u}_s = 1.32$.

substituting the source size with an effective source size, now no longer a constant,

$$\ell_s(z_\ell, z_s) = \left(\frac{\theta_\ell}{\theta_e}(A_{\min}) - 1\right)\ell_e. \tag{13}$$

## 3 RESULTS

The calculated $\tau$–$z$ correlations depend on six unknown parameters: $q_0$, $H_0/m_\ell$, $\tilde{u}_\ell$, $\tilde{u}_s$, $\tilde{\ell}_s$ and $C_\theta$. We will assume that $u_\ell = u_s = 500$ km/s ($\tilde{u}_\ell = \tilde{u}_s = 1.32$ for $v_o = 380$ km/s towards right ascension 11.2h, declination $-6°$ (Lubin et al 1985)). We will assume that the QSO optical-UV continuum is emitted from the inner 20 Schwarzschild radii of the accretion disk,

$$\ell_s = 6 \times 10^{14} M_8 \text{ cm}, \tag{14}$$

and

$$\tilde{\ell}_s \sim 0.01 \sqrt{\frac{h}{m_1}} M_8, \tag{15}$$

where $M_8$ is the black hole mass in $10^8 M_\odot$, $h = H_0/(100 \text{ km s}^{-1} \text{ Mpc}^{-1})$ and $m_1 = m_\ell/M_\odot$. $T$ is translated into an upper limit on $\tilde{\tau}$ with equation 9,

$$\tilde{\tau} = 0.016 \sqrt{\frac{h}{m_1}} \frac{\tau}{1 \text{ yr}}. \tag{16}$$

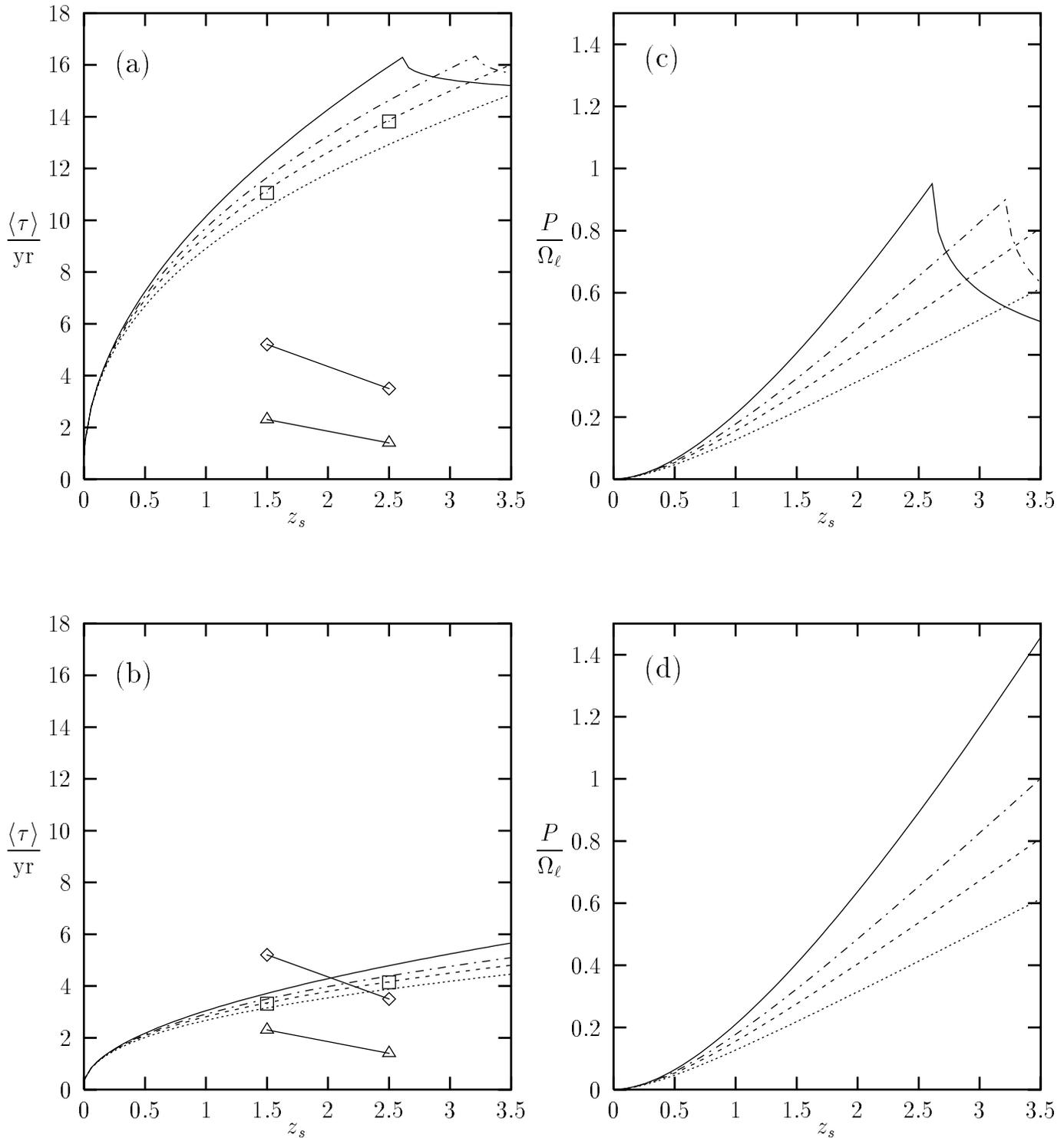

**Figure 4.** The $\tau$–$z$ correlations and lensing probability for the Hawkins sample ($\alpha = 125°$, $T = 17$ yr) with $\tilde{\ell}_s = 0.05$ (point-like source), $u_\ell = u_s = 500$ km/s and $C_\theta = 10$ for $q_0 = 0$ (solid line), $1/4$ (dash-dotted line), $1/2$ (dashed line) and $1$ (dotted line). The $q_0 = 1/2$ calculated timescale ($\square$), averaged over $z_s$ bins 1–2 and 2–3, are compared to the observed exponential fit timescale for the high ($\diamond$) and low ($\triangle$) luminosity bins. a) $\sqrt{h/m_1} = 3$ ($M_8 = 1.7$). b) $\sqrt{h/m_1} = 10$ ($M_8 = 0.5$). c, d) The specific lensing probability for (a) and (b), respectively.

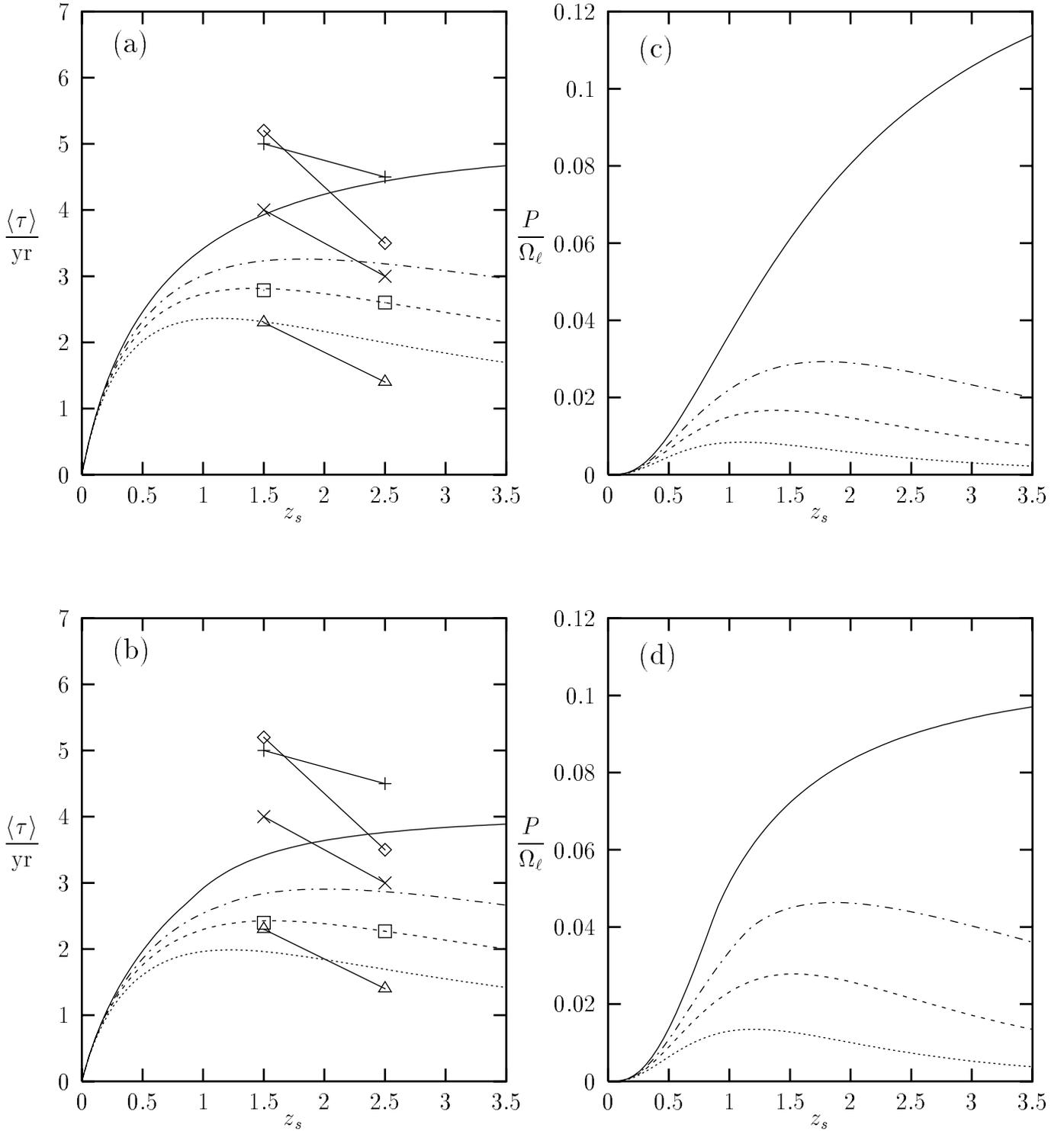

**Figure 5.** The $\tau$–$z$ correlations and lensing probability for the Hawkins sample ($\alpha = 125°$, $T = 17$ yr) for an extended source with $\sqrt{h/m_1} = 10$, $u_\ell = u_s = 500$ km/s. a) Approximate timescale, $\tilde{\ell}_s = 1$ ($M_8 = 10$), $C_\theta = 1$. b) Amplification threshold timescale, $\tilde{\ell}_s = 3$ ($M_8 = 30$), $A_{\min} = 1.1$. c, d) The specific lensing probability for (a) and (b), respectively. $q_0 = 0$ (solid line), 1/4 (dash-dotted line), 1/2 (dashed line) and 1 (dotted line). The $q_0 = 1/2$ calculated timescales ($\square$), averaged over $z_s$ bins 1–2 and 2–3, are compared to the observed exponential fit timescale and ACF zero crossing timescale for the high ($\diamond$,+ respectively) and low ($\triangle$,×) luminosity bins.

redshift. Consequently, the lensing probability is very small, $P/\Omega_\ell \sim 0.02$ (Fig. 5c). Long term QSO variability is known to be very common (e.g. Hawkins & Véron 1993), and therefore this value is unacceptably low even for $\Omega_\ell \sim 1$. This rules out the idea of microlensed extended sources as an explanation for the Hawkins correlations, unless one assumes that the *local* density of lenses within $\sim 300h^{-1}$ Mpc of the solar system is at least $\Omega_\ell \sim 10$.

Figures 5b and 5d show the lensing timescales and probabilities as calculated for $\tilde{\ell}_s = 3$ and an amplification threshold of $A_{\min} = 1.1$ (equation 13) with Schneider's (1993) analytical approximation of $A$ for a constant surface brightness distribution. The approximate results (figs. 5a and 5c) reproduce the exact ones quite well. The fact that the approximate method requires a smaller source size is to be expected since it tends to over-estimate the lensing time at the lower $z_\ell$ values, before the onset of the angular size ratio cutoff, $C_\theta$. The overall similarity in the results, even for an extended source, where the approximation is crudest (for a point-like source the differences are negligible), demonstrates that the approximate method is adequate for studying the general trends in the $\tau$–$z$ correlations and that these depend only weakly on the details of the source surface brightness.

The slopes of the the observed $\tau$–$z$ correlations can be made smaller, and closer to the calculated ones, if the timescales are defined to be the zero crossing point of the ACF. In this case, however, the timescales are increased to $\tau(\bar{z}_s = 1.5) \sim 4$, $\tau(\bar{z}_s = 2.5) \sim 3$ for the low luminosity bin and to $\sim 5$ and $\sim 4.5$ respectively for the high luminosity bin and are a factor of two higher than those calculated.

### 3.2 Positional and wavelength dependence of the $\tau$–$z$ correlations

If the typical velocities of lenses and sources are of the same order of magnitude as that of the solar system relative to the CMB, there may be detectable differences in the timescales of the $\tau$–$z$ correlations for small angular area surveys at different angles to $\vec{v}_o$. Figure 6 shows that the typical timescales for microlensing of QSO samples in the direction of the observer's velocity are longer than for those perpendicular to it and that the difference becomes more marked as $\tilde{\ell}_s$ increases. A factor of up to $\sim 1.2$ is obtained in the example of an extended source with small angular size cutoff, such that displays an inverse $\tau$–$z$ correlation. The positional dependence is not large, but if it is actually discovered, it may provide conclusive evidence for the microlensing hypothesis.

Another possible signature of microlensing is the different response of the light curve in different bands. In the framework of the accretion disk model, the high and low energy photons are emitted from the inner and outer parts of the disk, respectively, and therefore have different effective source sizes. Figure 7 shows the behavior of $\tilde{\tau}$ for two bands of fixed QSO rest-frame wavelength, schematically labeled as $B$ and $R$, with the $B$ band photons emitted from half the radius of the $R$ band photons. When the source size is small, the typical timescale of the amplification peaks in the $R$ band is longer than that of the $B$ band. However, for an extended source with a small angular ratio cutoff, the trend is reversed and the $R$ band light curve has significantly narrower amplification peaks since it is being amplified by nearer lenses. The mean time *between* the $R$ band peaks is, however, expected to be longer, since the lensing probability is smaller.

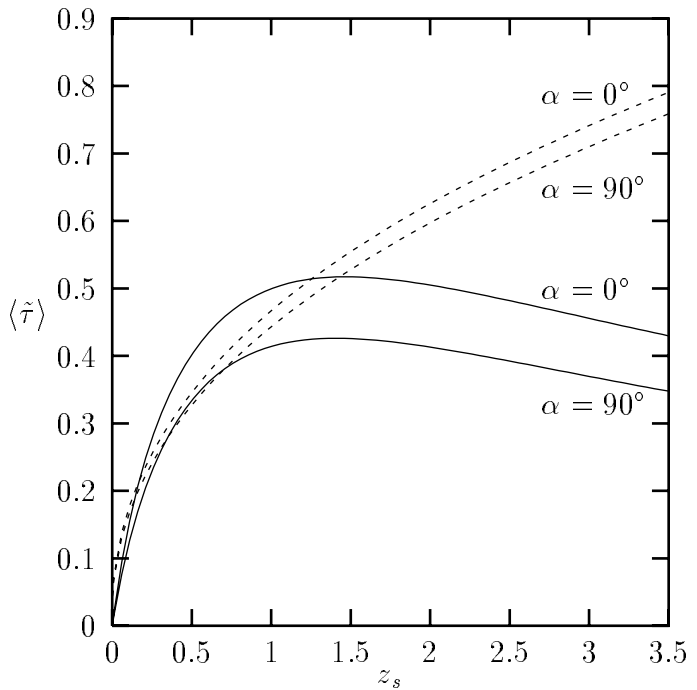

**Figure 6.** The dependence of the $\tau$–$z$ correlations on the survey angle, $\alpha$, for $q_0 = 1/2$, $\tilde{u}_\ell = \tilde{u}_s = 1.32$ and $C_\tau = \infty$. The case of $\alpha = 0°$ is compared to $\alpha = 90°$ for an extended source with $\tilde{\ell}_s = 1$ and $C_\theta = 1$ (solid line) and for a point-like source source with $\tilde{\ell}_s = 0.05$ and $C_\theta = 10$ (dashed line).

## 4 DISCUSSION AND SUMMARY

The $z_s$ dependence of lensing-induced variability reflects an interplay between the lens cross-section, lens density, the source angular size and the combined motion of the observer, lens and source on cosmological scales. Even with the assumption of a uniform lens population and QSO sample, there remain in the problem six unknown free parameters ($q_0$, $H_0/m_\ell$, $\tilde{u}_\ell$, $\tilde{u}_s$, $\tilde{\ell}_s$ and $C_\theta$), some of which are partly degenerate (e.g. a small source and low source and lens velocities can lead to similar lensing timescales as the opposite combination). The following discussion will therefore be confined to the general trends in the $\tau$–$z$ correlations and not to specific values of the parameters or to the bounds that can be set on them.

The $z_s$ dependence of microlensing variability is distinctly different from the $1 + z_s$ cosmological time dilation of intrinsic variability. Unlike the constant gradient of the intrinsic, dilated timescale, the lensing timescale gradient decreases as $z_s$ increases, and in some cases even becomes negative (e.g. figures 4a and 5a). This trend appeared in all of the many parameter combinations that were checked.

The observed $\tau$–$z$ correlations can, in some situations,

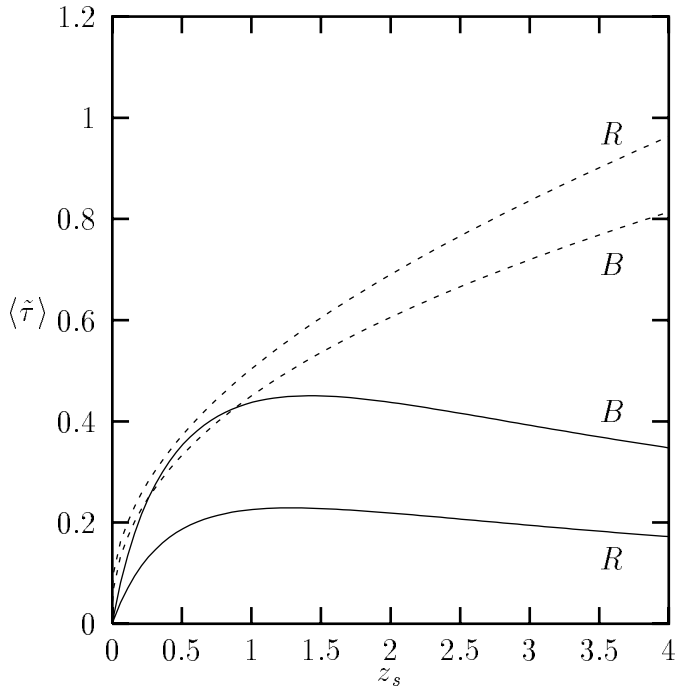

**Figure 7.** The dependence of the $\tau$–$z$ correlations on the rest-frame wavelength band for $q_0 = 1/2$, $\tilde{u}_\ell = \tilde{u}_s = 1.32$ and $C_\tau = \infty$. The $B$ band photons are assumed to be emitted from half the radius of the $R$ band photons. The timescales in the $B$ and $R$ bands are compared for an extended source of $\tilde{\ell}_s = 1$ and 2, respectively, and $C_\theta = 1$ (solid line) and for a point source of $\tilde{\ell}_s = 0.05$ and 0.1, respectively, and $C_\theta = 10$ (dashed line).

depend strongly on the observational procedure and its limitations: the finite duration of the monitoring campaign and the finite photometric sensitivity. When the lightcurves display clearly distinguishable peaks, there are two typical timescales: the mean peak width and the mean time between peaks. The results presented here apply to the peak width only. It is important that the method chosen to quantify the timescale of the lightcurve (e.g. ACF zero peak width, mean FWHM or power spectrum maximum) be matched to the definition of the timescales used in the theoretical predictions and that these take into account the observational limitations. Failure to do so may lead to spurious relationships between the predictions and the data.

The closest resemblance between the Hawkins results and the models investigated here is obtained for an extended source monitored with a moderate photometric sensitivity ($\Delta m \sim 0.1$m) and a $q_0 \geq 1/4$ cosmological model. The calculated timescales decrease with $z_s$, as do the observed ones, but not as steeply (the fit was obtained by trial and error and it is possible that a better one exists). In this case the lenses that are causing the variability are within $\sim 300 h^{-1}$ Mpc of the solar system and the lensing probability is very low unless the local lens density within this radius is very high (equivalent to $\Omega_\ell \sim 10$). The wavelength dependence of the $\tau$–$z$ correlations can be used to test this hypothesis independently of arguments about the likelihood of having an atypically high local lens density. In this case the mean peak width in the longer wavelength is expected show *shorter* timescales, contrary to the intuitive expectation. The sample position dependence can be also used for this purpose, although the decrease of the timescales with $\alpha$ may be harder to detect.

Microlensing of point-like sources (i.e. point-like with respect to the typical lens mass) can be attained for $\Omega_\ell$ of the order of 1 or less. The $\tau$–$z$ correlations in this case are expected to rise monotonically (Fig. 4b), unlike the Hawkins results. This result is in agreement with that of Canizares (1982). The wavelength and positional dependence of the $\tau$–$z$ correlations are not very useful for identifying this case of microlensing. The shorter wavelengths are predicted to vary on shorter timescales, as expected also for intrinsic variability, and the positional dependence may very well be undetectable.

Many assumptions are involved in calculating the $\tau$–$z$ correlations. The most problematic is that of a uniform QSO sample, as QSO are known to undergo a significant evolution. The QSO analyzed by Hawkins were binned by absolute magnitude, and this restricts the possible variance in the properties of objects in the same bin. Nevertheless, it is conceivable that QSO of similar luminosities but of different cosmological epochs may have different source sizes. Dealing with such effects requires detailed modeling of the source evolution and emission mechanism. An additional $z_s$ dependence of $\ell_s$ can be introduced by a temperature gradient in the source, such as is expected in accretion disk models. Observations in a given filter detect different parts of the source at each redshift, depending on the source radial temperature profile. This effect was studied in the specific case of a thin accretion disk, where the surface temperature decreases as $\ell_s^{-3/4}$ (e.g. Shapiro & Teukolsky 1983) and therefore $\ell_s \propto (1 + z_s)^{-4/3}$. It is found that this does not change the qualitative results in the case of point-like sources. For extended sources, the decreasing effective source size leads to $\tau$–$z$ correlations that resemble those of the point sources shown in figure 4b. Thus, a temperature gradient increases the discrepancy between the theoretical prediction and the Hawkins results. The wavelength dependence of extended sources is also affected by the temperature gradient when the $R$ and $B$ bands are fixed filters in the observer's frame. For the same parameters as in figure 7, the $R$ band varies faster than the $B$ band for $z_s < 2.5$, but this trend reverses at higher redshifts and resembles that of point-like sources.

Another potential problem is that the definition of lensing timescale is based on the assumption of a compact lens, whose properties depend only on its mass and is characterized by a symmetric lightcurve. This is no longer true for caustic networks where the lensing is a collective effect of many compact objects, which depends also on their distribution and can give rise to complicated amplification behavior (e.g. Schneider & Weiss 1993). Finally, it is unclear whether comparing the calculated $\tau$–$z$ correlations to an exponential fit of the empirical ACF zero-lag peak is theoretically justified. A more cautious approach would be to fit the empirical ACF to a calculated one, which should also take into account overlapping amplification peaks, rather than just to a mean timescale.

It is concluded that the calculated $\tau$–$z$ correlations do not support the idea that the Hawkins results can be attributed to the microlensing of a uniform QSO sample by a uniform distribution of lenses. The nature of the observed $\tau$–$z$ correlations may reflect the details of the observational techniques employed no less than the underlying cosmological model.

## Appendix

This appendix summarizes the identities required for expressing the distances, timescales and probabilities of the microlensing in terms of the cosmological model parameters, the lens and source redshifts and the lens mass.

The constant comoving lens density, the light path interval and the volume element are given by

$$\rho_\ell = \rho_0(1+z)^3, \tag{A1}$$

$$c|dt| = R_\mathrm{H} \frac{dz}{(1+z)^2\sqrt{1+2q_0 z}}, \tag{A2}$$

and

$$dV = R_\mathrm{H}^3 \frac{[q_0 z + (q_0-1)(\sqrt{1+2q_0 z}-1)]^2}{q_0^4(1+z)^6\sqrt{1+2q_0 z}} dz\, d\Omega_s, \quad q_0 > 0 \tag{A3}$$

$$dV = R_\mathrm{H}^3 \frac{z^2(2+z)^2}{4(1+z)^6} dz\, d\Omega_s, \quad q_0 = 0. \tag{A4}$$

where $d\Omega_s = \pi \ell_s^2/D_{os}^2$ is the solid angle subtended by the source and $R_\mathrm{H} = c/H_0$.

The Einstein radius $\ell_e$ is defined as

$$\ell_e^2 = \frac{4Gm_\ell}{c^2} \frac{D_{o\ell} D_{\ell s}}{D_{os}} \tag{A5}$$

where $D_{12}$ is the angular diameter distance between points 1 and 2 ($z_1 < z_2$), given by

$$D_{12} = R_\mathrm{H} \frac{(1-2q_0)(G_1-G_2) + (G_1 G_2^2 - G_1^2 G_2)}{2q_0^2(1+z_1)(1+z_2)^2}, \quad q_0 > 0 \tag{A6}$$

$$D_{12} = R_\mathrm{H} \frac{(z_2-z_1)(2+z_1+z_2)}{2(1+z_1)(1+z_2)^2}, \quad q_0 = 0. \tag{A7}$$

with $G_i = (1+2q_0 z_i)^{1/2}$ (Blandford & Kochanek 1987).

The coordinate distance is given by

$$r = \frac{R_\mathrm{H}}{R_0} \frac{[q_0 z + (q_0-1)(\sqrt{1+2q_0 z}-1)]}{q_0^2(1+z)}, \quad q_0 > 0 \tag{A8}$$

$$r = \frac{R_\mathrm{H}}{R_0} \frac{z(2+z)}{2(1+z)}, \quad q_0 = 0, \tag{A9}$$

where $R_0$ is the present-day expansion factor.

The lens path length, $\ell_\ell$, is given by

$$\ell_\ell = R_\ell r_\ell(\theta_s + \theta_e) = \ell_e + \frac{D_{o\ell}}{D_{os}} \ell_s. \tag{A10}$$

The observer's path length $\ell_o$ is given, to 1st order in $\epsilon$, by

$$\ell_o = R_0 r_0 \epsilon = (1+z_\ell) \ell_e \frac{r_0}{x_\ell}. \tag{A11}$$

$r_0$ and $x_\ell$ are the solutions of the following equations (see e.g. Weinberg 1972)

$$\begin{align}
\ell_e &= R_\ell x_\ell \epsilon, \\
\ell_s &= R_s x_s \epsilon, \\
r_\ell &= r_0 \sqrt{1-kx_\ell^2} - x_\ell \sqrt{1-kr_0^2}, \\
r_s &= r_0 \sqrt{1-kx_s^2} - x_s \sqrt{1-kr_0^2}, \tag{A12}
\end{align}$$

where the radial coordinate systems $r$ and $x$ are defined in figure 1b, $R_s$ is the scale factor at the source emission and $k$ is the constant of curvature, which in a non-relativistic, matter dominated Friedmann universe is given by

$$k = \left(\frac{R_0}{R_\mathrm{H}}\right)^2 (2q_0-1). \tag{A13}$$

In terms of the normalized radial coordinates, $\tilde{r} = (R_0/R_\mathrm{H})r$ and $\tilde{x} = (R_0/R_\mathrm{H})x$, the solution of equations A12 for $x_\ell$ is given by the root of

$$\begin{align}
0 = &[4k(\tilde{r}_\ell + \beta\tilde{r}_s)(\tilde{r}_s + \beta\tilde{r}_\ell)\beta + (\beta^2-1)^2]\tilde{x}_\ell^4 - \\
&[4(\tilde{r}_\ell + \beta\tilde{r}_s)^2 - 2(\tilde{r}_s^2 - \tilde{r}_\ell^2)(\beta^2-1) - \\
&4\tilde{r}_\ell \tilde{r}_s(\tilde{r}_\ell + \beta\tilde{r}_s)(\tilde{r}_s + \beta\tilde{r}_\ell)k]\tilde{x}_\ell^2 + \\
&[\tilde{r}_s^2 - \tilde{r}_\ell^2]^2, \tag{A14}
\end{align}$$

with the parameter $\beta$ defined as

$$\beta = \frac{(1+z_s)}{(1+z_\ell)} \frac{\tilde{\ell}_s}{\tilde{\ell}_e}. \tag{A15}$$

## Acknowledgements

It is a pleasure to thank Dani Maoz for his advice and encouragement. I am grateful to Hagai Netzer and Amiel Sternberg for their helpful comments on the final draft. Partial financial support by the US-Israel Binational Science Foundation grant no. 89-00179 is acknowledged.